\documentclass[%
 reprint,
 amsmath,amssymb,
 aps,
 prl,
]{revtex4-2}

\usepackage{graphicx}
\usepackage{dcolumn}
\usepackage{bm}

\usepackage{xcolor}

\begin{document}


\title{Dynamics of Soap Bubble Inflation}

\author{Saini Jatin Rao}

\author{Siddhant Jain}

 \author{Saptarshi Basu}
 \altaffiliation[Also at ]{Interdisciplinary Centre of Energy Sciences, Indian Institute of Science, Bengaluru, 560012, India.}

 \email{sbasu@iisc.ac.in}
  
\affiliation{%
 Department of Mechanical Engineering, Indian Institute of Science, Bengaluru, 560012, India
}%

\date{\today}

\begin{abstract}

Bubbles have always captivated our curiosity with their aesthetics and complexities alike. While the act of blowing bubbles is familiar to everyone, the underlying physics of these fleeting spheres often eludes reasoning. In this letter, we discuss the dynamics of inflating a soap bubble using controlled airflow through a film-coated nozzle. We assess and predict the rate of inflation by varying the source pressure. Visualising the previously unexplored internal flow reveals that air enters the bubble as a round jet, emerging from the nozzle opening and impinges on the expanding concave bubble interface to form a toroidal vortex. Several scaling laws of the associated vortical flow spanning the entire bubble and the vortex core are reported. The observed dynamics of this bubble-confined vortex ring formation indicate universality in certain aspects when compared to the free laminar vortex rings.

\end{abstract}

\maketitle

\begin{figure}[hb]
\includegraphics[width=1\linewidth]{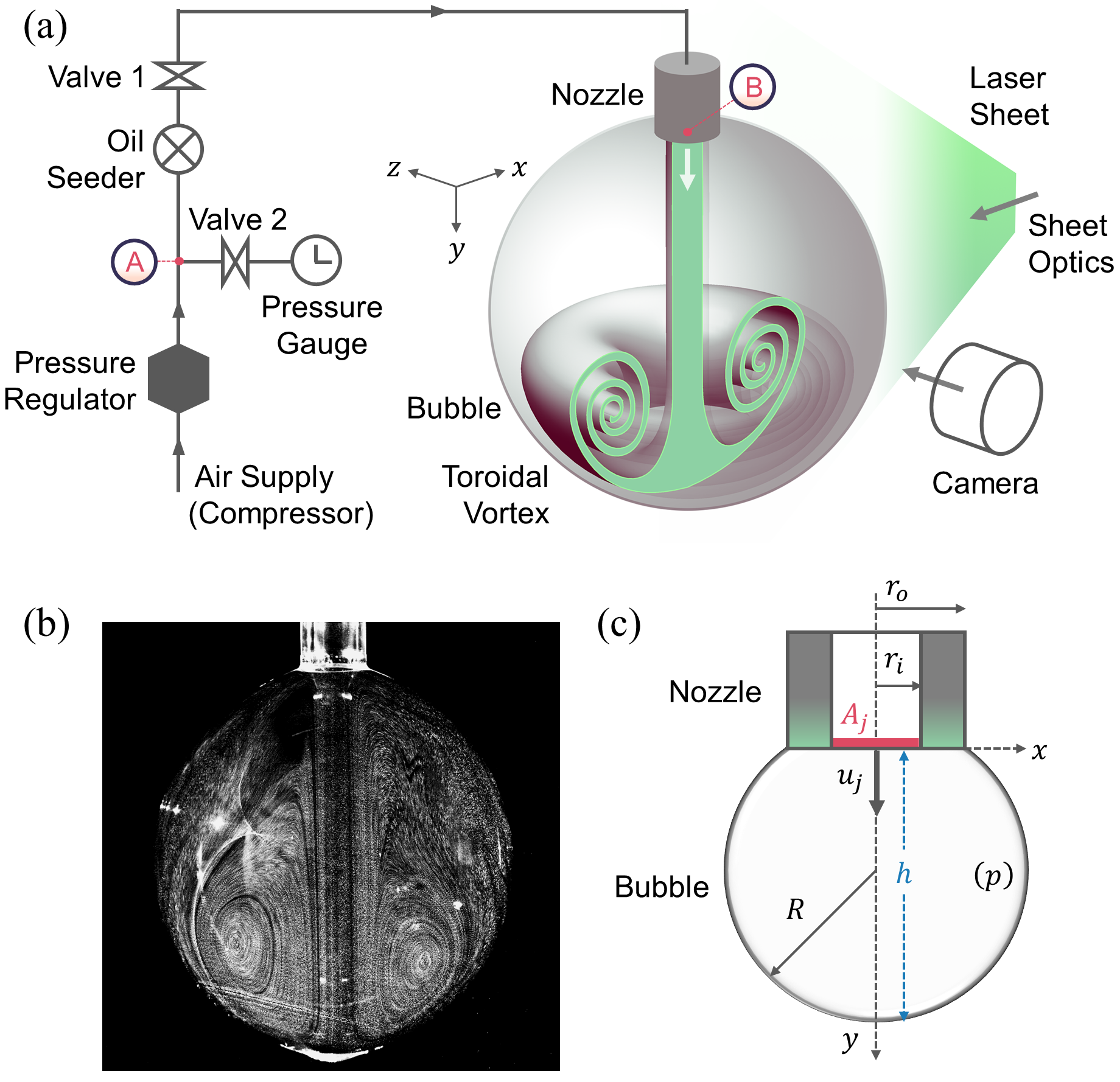}
\caption{\label{fig:setup} (a) A schematic of the experimental setup for inflating a soap bubble with an air supply at a prescribed pressure, seeded with olive oil droplets. A laser sheet is used to illuminate these particles for flow visualisation captured through a high-speed camera. (b) Long exposure image of the flow within the bubble during inflation depicting vortex structure (c) Nomenclature associated with the nozzle (inner radius $r_i$, outer radius $r_0$, jet area $A_j$), bubble geometry (spherical cap radius $R$, height $h$) and airflow (jet velocity $u_j$, static gauge pressure inside bubble $p = 4\sigma/R$ where $\sigma$ is air-soap interface surface tension). }
\end{figure}

Soap bubbles have been a source of wonder for researchers and the general public alike for generations, yet our continued exploration still unveils nuances that keep us engaged with these iridescent objects \cite{boys_soap_1959}. Among these exciting features are the vivid interference patterns \cite{lalli_coherence_2023}, bursting films \cite{lhuissier_soap_2009}, minimal surfaces \cite{giomi_minimal_2012} and even the demonstration of a soap bubble as an optical cavity to generate a laser \cite{korenjak_smectic_2024}. This paper specifically delves into the physics of soap bubble inflation, revealing the remarkable flow that happens inside. Notably, the internal flow exhibits a toroidal vortex structure as illustrated in Fig.~\ref{fig:setup}.

The dynamics associated with bubble inflation have been examined from different perspectives throughout the literature. In the most general sense, the soap bubble anatomy consists of a thin film of soap liquid stretching against a gas flow. This soap film is usually pinned or held over an aperture through which the air is introduced, enabling the inflation process. The aperture can be a ring or a wand, where a free jet or stream of gas is used for blowing the bubble \cite{salkin_generating_2016, zhou_formation_2017, hamlett_blowing_2021, ganedi_equilibrium_2018}. Another possibility is a closed configuration where an opening at the tube end with the gas efflux is the aperture over which the soap film is pinned \cite{grosjean_unstable_2023}. These studies primarily focused on the motion or stretching of the soap film. It has been shown that in the case of blowing using a free gas jet, there exists a critical velocity beyond which a bubble forms and detaches from a soap film \cite{salkin_generating_2016, zhou_formation_2017}. Blowing a bubble through a constriction in a fixed mass system depicts an unstable growth of the bubble interface \cite{grosjean_unstable_2023}. However, beyond these considerations, the associated internal flow structure of the gas phase being used for inflation is rarely explored or reported.  

\begin{figure*} [ht]
\includegraphics[width=0.95\linewidth]{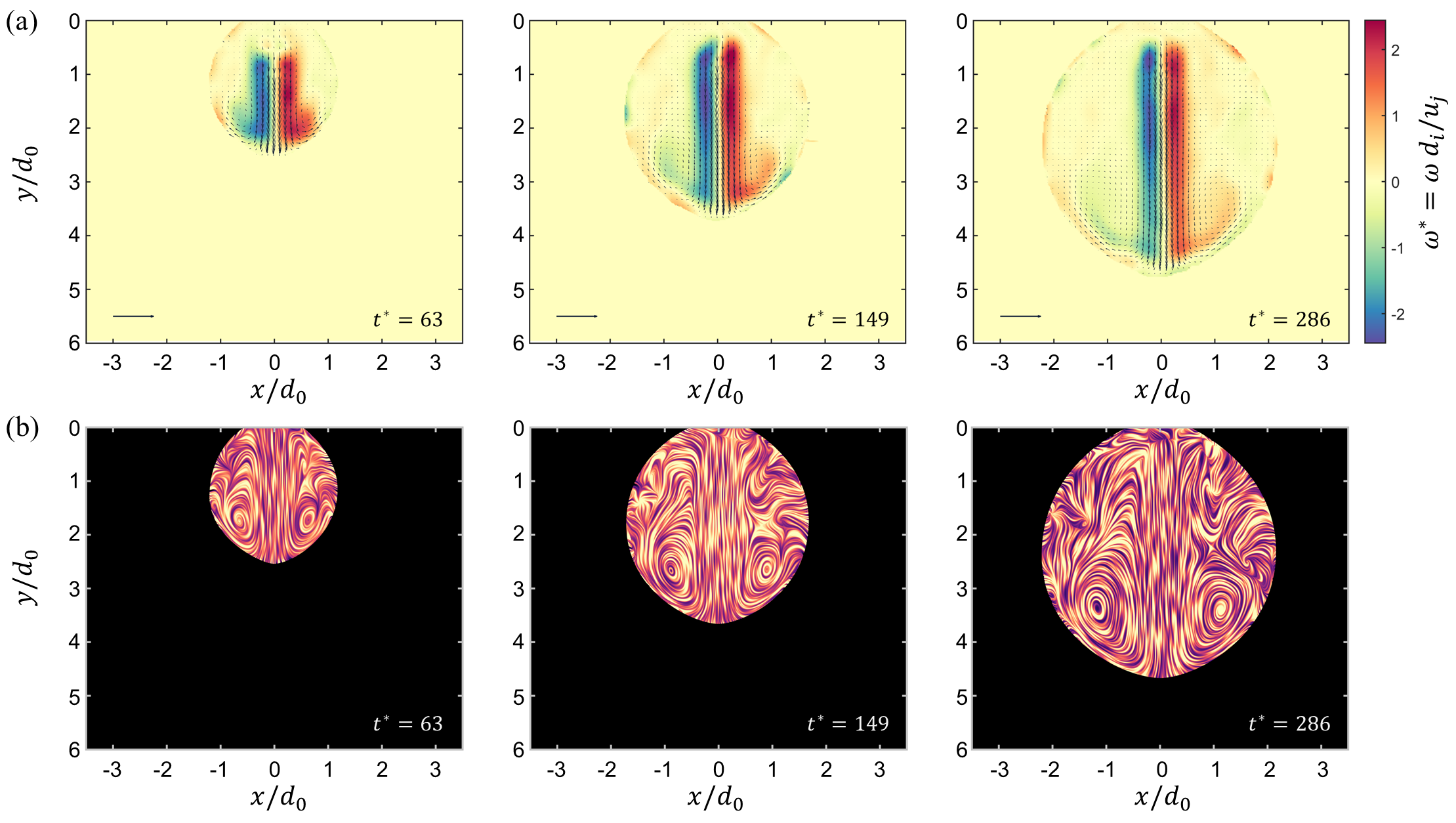}
\caption{\label{fig:maps} (a) Vorticity and vector fields associated with internal flow of an inflating bubble at various normalised time instances for the case of dimensionless source pressure ${\widetilde{p}}_0 \approx 28$ (b) LIC (Line Integral Convolution) flow visualisation for the same flow instances.}
\end{figure*}

The soap films were also extended to study two-dimensional flows where the incumbent vortical dynamics were rigorously observed and studied for the flowing film \cite{vorobieff_soap_1999, roushan_universal_2005} and bubbles \cite{seychelles_thermal_2008,meuel_intensity_2013}. This study, however, focuses on the vortical dynamics of the gas phase flow within a growing soap bubble.

The anatomy of an inflating bubble can be extended to other systems where a membrane or interface is stretched against a fluid flow. This involves problems spanning from filling a simple rubber balloon \cite{ilssar_inflation_2020} to the flow focusing within microdroplets \cite{vagner_vortex_2021}. Containers like plastic bottles or glass vessels are produced by blow moulding, i.e. inflating a plug in a mould of a prescribed shape, where flow-induced cooling depends on the internal fluid motion \cite{zimmer_experimental_2015}. Soft robotics and medical devices \cite{manfredi_soft_2019} involve controlled inflation and deflation of tubes, balloons or chambers, where a uniform flow and pressure distribution might be necessary. Furthermore, the membrane-based systems involving engulfment flows induced by forced membrane motion will also depict similar flow characteristics. The jellyfish-like locomotion \cite{costello_hydrodynamics_2021, baskaran_lagrangian_2022}, the flow inside the heart chambers \cite{di_labbio_material_2018} or even artificial organs are a few examples.

For the aforementioned systems, the precise knowledge of internal flow dynamics of the fluid phase enclosed within a membrane-like entity is deemed necessary for a better understanding and control of the associated phenomenon. The exact flow conditions will differ vastly from our canonical bubble inflation problem; however, a prominent vortical structure is usually associated, either due to fluid-structure interaction or unsteady inlet flow conditions. This problem thus casts a basis for a class of problems involving a confined, unsteady toroidal vortex continually being fed with a mass flux, unlike the traditional self-propelled free vortex rings, where the feeding is discontinued after it pinches off from the orifice \cite{gharib_universal_1998, krieg_new_2021}.

\begin{figure}[h]
\includegraphics[width=0.7\linewidth]{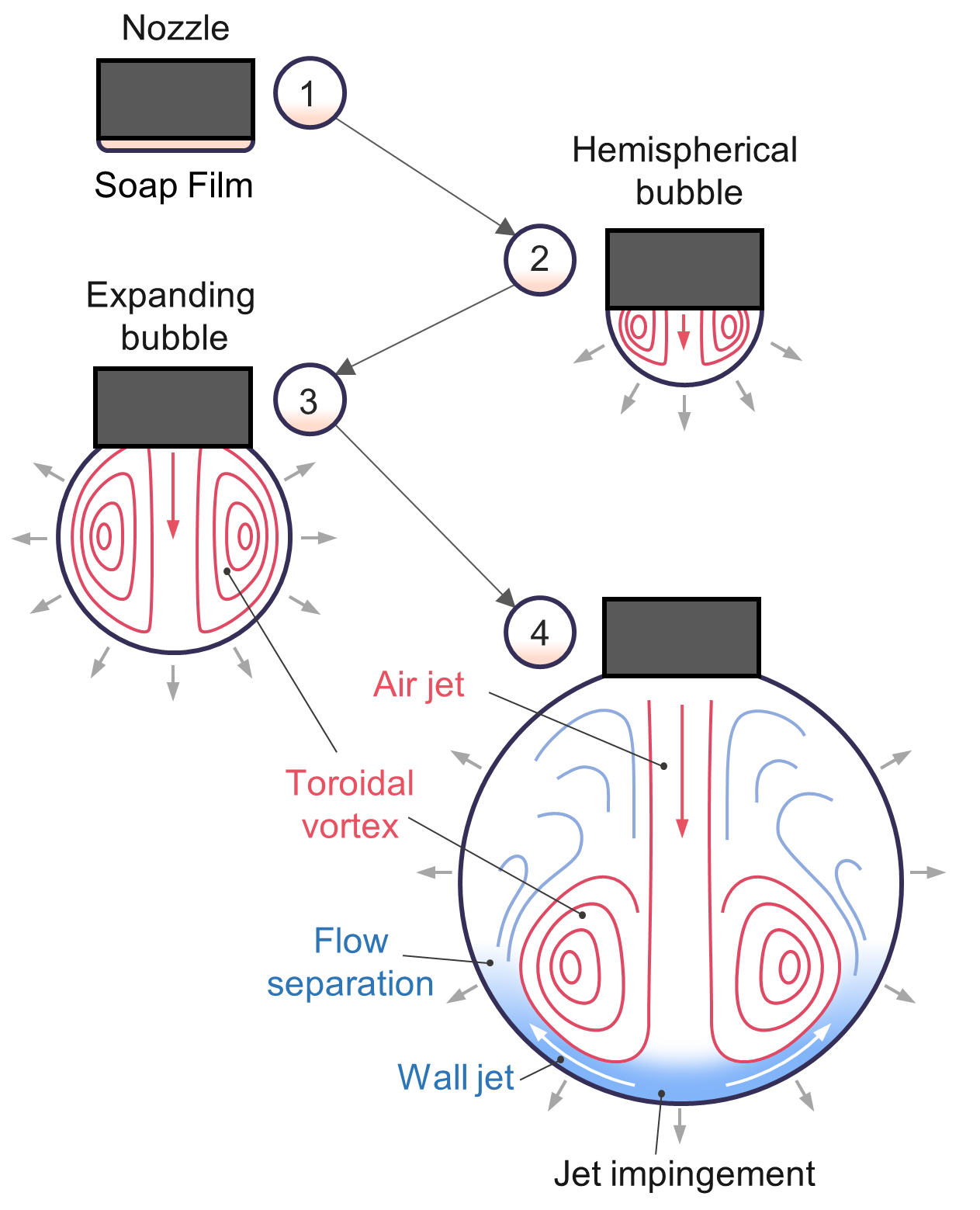}
\caption{\label{fig:schematic} A schematic depicting the sequence of events for a soap bubble inflation (1) Thin film deposited over the nozzle aperture (2) Film expansion induced by airflow leading to hemispherical bubble (3) Bubble growth conforming a spherical cap with the gas phase flow within depicting a toroidal vortex spanning the whole volume (4) Further film expansion with the incoming round air jet impinging on bubble interior, developing a wall jet which separates to form a localised vortex.}
\end{figure}

In the current exposition, we consider inflating a soap bubble by expanding a soap film deposited at the tube end opening with air. The first part of the study looks into the dynamics associated with the soap film motion and uses first principles to deduce the scaling laws for the bubble growth rate. The second part delves further into the internal airflow dynamics within this growing bubble. The incoming airflow conforms into a toroidal vortex initially. Eventually, as the bubble grows, air enters as a round jet which impinges on the concave bubble surface, forming a wall jet that eventually separates to form a localized toroidal vortical structure (See Fig.~\ref{fig:maps}, \ref{fig:schematic} and supplemental movies 1-6). 


\vspace{3mm}
\textbf{Experiments:}
The schematic of the experimental setup is depicted in Fig.~\ref{fig:setup}a. Air supply at a constant pressure from the reservoir is utilised for bubble inflation, where the downstream pressure is controlled using a pressure regulator. A seeder arrangement in line with the flow circuit introduces olive oil droplets of size $\sim 1-5 \mu m$, which is essential for internal airflow visualisation during bubble blowing. A manometric gauge is used for pressure measurement between the regulator and the seeder. Through valve '1' and PU tubes, the output of this apparatus is connected to an acrylic nozzle at the other end having inner and outer diameters of $d_i = 2r_i = 6mm$ and $d_0 = 2r_0 = 10mm$, respectively. To generate a soap bubble, the nozzle is dipped in a soap solution (Commercial - Bubble Magic) to deposit a soap film at the opening. For repeatability, the nozzle tip is dipped to approximately the same depth of $\sim 1-2mm$ for each run, and the nozzle end is held within a large acrylic enclosure. Valve '2' is closed, and valve '1' is opened during the inflation process, where the input pressure is pre-adjusted through the regulator. To measure the actual stagnation pressure just after the regulator, valve '1' is closed, and valve '2' is opened after every run. The pressure measured through the manometer is then considered an equivalent source supply pressure for the system. The air-liquid surface tension $\sigma = 28.56 mN/m$ of the soap solution was measured using the pendant droplet method, and viscosity $\eta_{\mathrm l} = 132.49 mPa\ s$ was determined using an Anton Paar model rheometer.

The flow visualisation is achieved using a high-speed dual pulsed Nd:YLF laser (pulse energy of 30 mJ, emission wavelength 527 nm, Photonics Inc.), where a laser sheet from a cylindrical lens is passed through the symmetry axis of the bubble and tube system. The images of the illuminated seeded particles for particle image velocimetry (PIV) were captured at 95 frames per second, using a high-speed camera (Photron Mini UX) at a pixel resolution of $1280\times1024$ pixels with a field of view of $86\times70 mm$. The outlines of the bubble interfaces in these images (See Fig.~\ref{fig:setup}b) were used to determine the geometrical parameters like height $h$ and radius $R$, considering the bubble to be a spherical cap pinned over the outer edge of the nozzle. The associated nomenclature is depicted in Fig.~\ref{fig:setup}c.


For PIV, the images were post-processed in Davis 8.4 software to obtain the velocity field of the internal flow. This involves a cross-correlation technique with a constant multi-pass interrogation window of size $32\times32$ pixels with $50\%$ overlap. The velocity vectors and the vorticity field for a particular source pressure at different normalised time instances are depicted in Fig.~\ref{fig:maps}a (These normalised parameters are defined later). The Line Integral Convolution (LIC) representation \cite{cabral_imaging_1993} depicting streamline-like features in a continuous fashion is illustrated in Fig.~\ref{fig:maps}b. Two counter-rotating vortices are clearly visible in Fig.~\ref{fig:maps}, portraying a dominant toroidal vortex engulfed within the bubble. The internal flowfield hence obtained will be utilised to assess the vorticity dynamics.

A phenomenological sequence of events for the inflation process is depicted in Fig.~\ref{fig:schematic}. Bubble inflation is achieved from the tube end with a soap film deposited at the opening. Initially, the incoming air fills the bubble volume, and the flow conforms to the concave confinement within the expanding film. The bubble is eventually hemispherical with a minimum radius of curvature and grows in radius later with further inflation. The internal flow forms a toroidal vortex, which spans the small bubble volume initially. Later, as the bubble grows, the incoming air assumes a round jet form. This laminar jet impinges the soap film at the other end of the bubble, forming a wall jet that moves along the curved inner surface of the bubble. This wall jet eventually separates to form a toroidal vortical structure engulfed within a growing bubble.

\begin{figure}[h]
\includegraphics[width=1\linewidth]{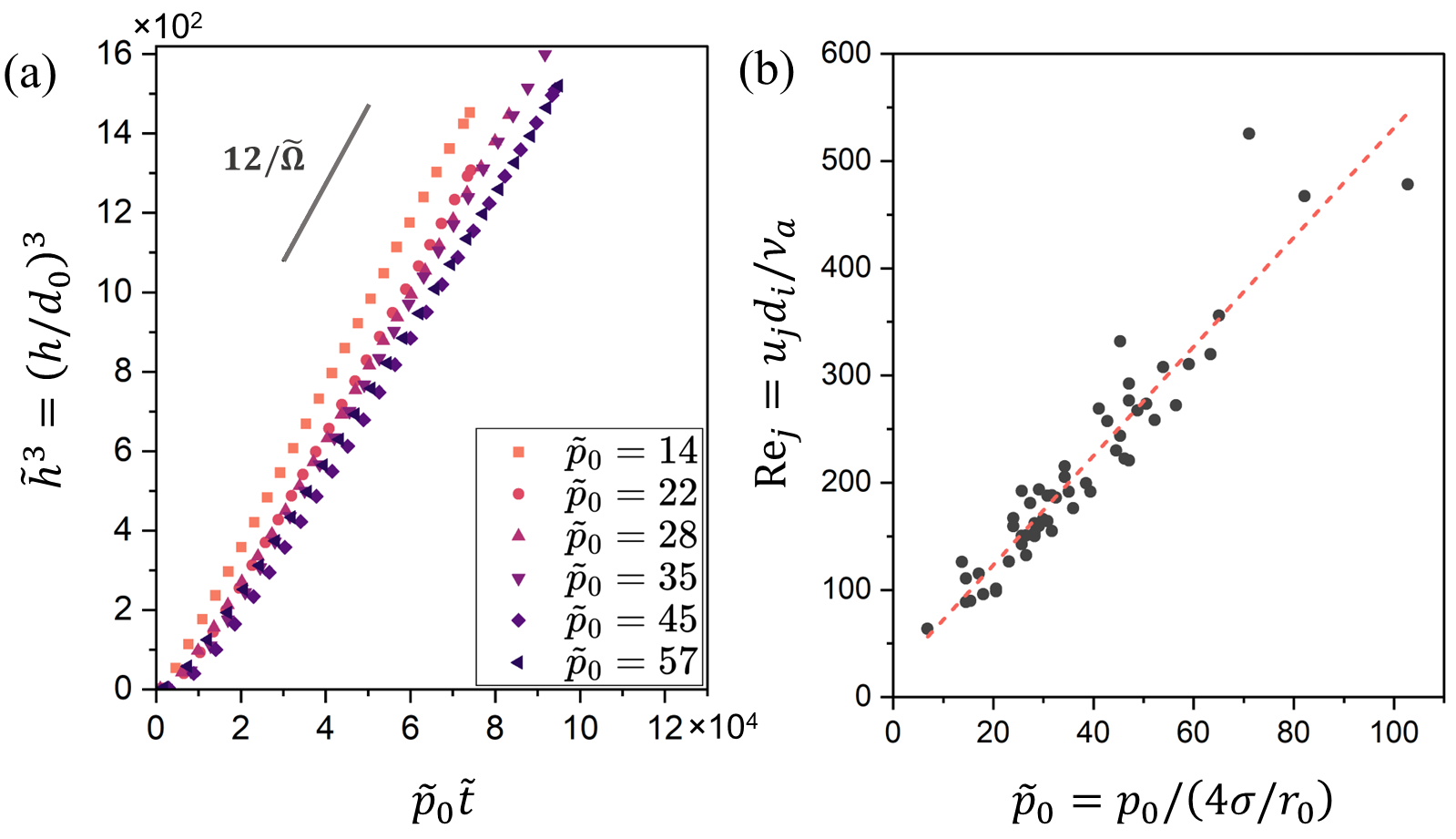}
\caption{\label{fig:inflation} (a) Growth of normalised bubble height ${\widetilde{h}}^3$ with a modified dimensionless time $\widetilde{p}_0 \widetilde{t}$ taking into account the source pressure, indicating linear variation of bubble volume with time.  (b) Steady-state air jet Reynolds number depicting a linear correlation with the dimensionless source pressure}
\end{figure}

\vspace{3mm}
\textbf{Inflation Dynamics:}
To emulate bubble inflation, a simple model based on Bernoulli's principle is employed. Applying this between points (A) and (B) (refer Fig.~\ref{fig:setup}a) gives a simple equation for airflow from the source to the bubble as
\begin{equation}  \label{eq:bernoulli}
    p_0 = \frac{4\sigma}{R} + \frac{1}{2} \rho_a u_j^2 + \Delta p_{\mathrm{loss}}
\end{equation}
where $R$ is the radius of the bubble considered to be as the spherical cap, $p_0$ is the stagnation gauge pressure at source (measured at (A) in Fig.~\ref{fig:setup}a), $\rho_a$ is the air density and $\sigma$ is the air-soap interface surface tension. The term $4\sigma/R$ represents the Laplace pressure and is assumed to be uniform static pressure within the bubble, including the nozzle exit (B) (see Fig.~\ref{fig:setup}a). The average air jet velocity $u_j$ at the nozzle exit can be expressed in terms of the volumetric flow rate $Q$ and exit area $A_j$ as $u_j = Q/A_j$. $\Delta p_{\mathrm{loss}}$ is the viscous pressure loss along the pipe and can be expressed as $\Delta p_{\mathrm{loss}} = \Omega Q$ where $\Omega$ is the equivalent frictional resistance for the piping system. For a given setup, the experimental parameter $\Omega$ should be roughly the same. For a Poiseuille's pipe flow $\Omega  \sim \frac{8\pi\mu_a L_{\mathrm{eff}}}{A_{\mathrm{eff}}^2}$, where $\mu_a$ is the air viscosity; $L_{\mathrm{eff}}$ and $A_{\mathrm{eff}}$ are the effective length and cross-sectional area for the piping system respectively. Eq.~\ref{eq:bernoulli} can be further simplified to obtain an expression for $Q$, which can be equated to the rate of increase in the bubble volume using mass balance. On simplification, the expression for the dimensionless version of the bubble height $\widetilde{h}$ (See Fig.~\ref{fig:setup}c for bubble and nozzle geometry details) is obtained as
\begin{equation}  \label{eq:inflation}
\frac{d\widetilde{h}}{d\widetilde{t}}=\frac{1}{1+{\widetilde{h}}^2}\left\{-\widetilde{\Omega}+\sqrt{{\widetilde{\Omega}}^2+8\left({\widetilde{p}}_0-\frac{2\widetilde{h}}{1+{\widetilde{h}}^2}\right)}\right\}
\end{equation}
\noindent where $\widetilde{h}=\frac{h}{r_o}$, $\widetilde{t}=\frac{t}{\tau}$ with time-scale $\tau=\frac{\pi}{2}\sqrt{\frac{\rho_ar_o^7}{A_j^2\sigma}}$, $\widetilde{\Omega}=\Omega\sqrt{\frac{A_j^2r_o}{\rho_a\sigma}}$ and ${\widetilde{p}}_0=\frac{p_0}{4\sigma/r_0}$. In the experiments, we observe a linear growth of the bubble volume with time, except for a very short early phase. In the limit of a large bubble $\widetilde{h}\gg1$, $h\approx2R$ and Eq.~\ref{eq:inflation} approximates to 
\begin{equation}  \label{eq:approx}
{\widetilde{h}}^3\approx\frac{12}{\widetilde{\Omega}} {\widetilde{p}}_0 \widetilde{t}
\end{equation}
The experimental data considered in the dimensionless form is hence plotted in Fig.~\ref{fig:inflation}a. ${\widetilde{h}}^3$ varies linearly with a modified dimensionless time $\widetilde{p}_0 \widetilde{t}$ taking into account the source pressure, as expected from Eq.~\ref{eq:approx}. The curves overlap and the slopes at various pressures assimilate to give a fairly constant $\widetilde{\Omega} \approx 600$, validating our model. Furthermore, as the bubble volume $\sim h^3$, the slope $dh^3/dt \propto Q \propto u_j$. The rate of the bubble volume growth was therefore used to evaluate the air jet velocity, which is unsteady only in the early phase of inflation. A steady state is approached as the bubble grows, i.e. the radius $R$ increases and the opposing Laplace pressure within the bubble diminishes. Defining the air jet Reynolds number as ${Re}_j = u_j d_i/\nu_a$, from Eq.~\ref{eq:approx}, ${Re}_j \propto {\widetilde{p}}_0$. The linear correlation is observed in the experimental realisations as depicted in Fig.~\ref{fig:inflation}b. From this analysis and experimental observations, we deduce the scaling for the bubble radius with time as $h \sim R \sim t^{1/3}$ for inflation using a constant pressure supply.

\vspace{3mm}
\textbf{Vortical Dynamics:}
The flow inside the bubble depicts a fascinating toroidal vortex structure as depicted in Fig~\ref{fig:maps}. The vortical flow spans the whole bubble volume and is similar to a Hill's vortex \cite{hill_vi_1997} in the early phase when the bubble is small as depicted in Fig~\ref{fig:schematic}. Impingement of the incoming round jet later forms a wall jet at the expanding concave interface, which separates eventually and rolls up to form a vortex localised to the lower part of the bubble as it grows (see Fig~\ref{fig:maps} and \ref{fig:schematic}). The incoming jet feeds mass to this growing vortex and the bubble as a whole. 

\begin{figure}[h]
\includegraphics[width=1\linewidth]{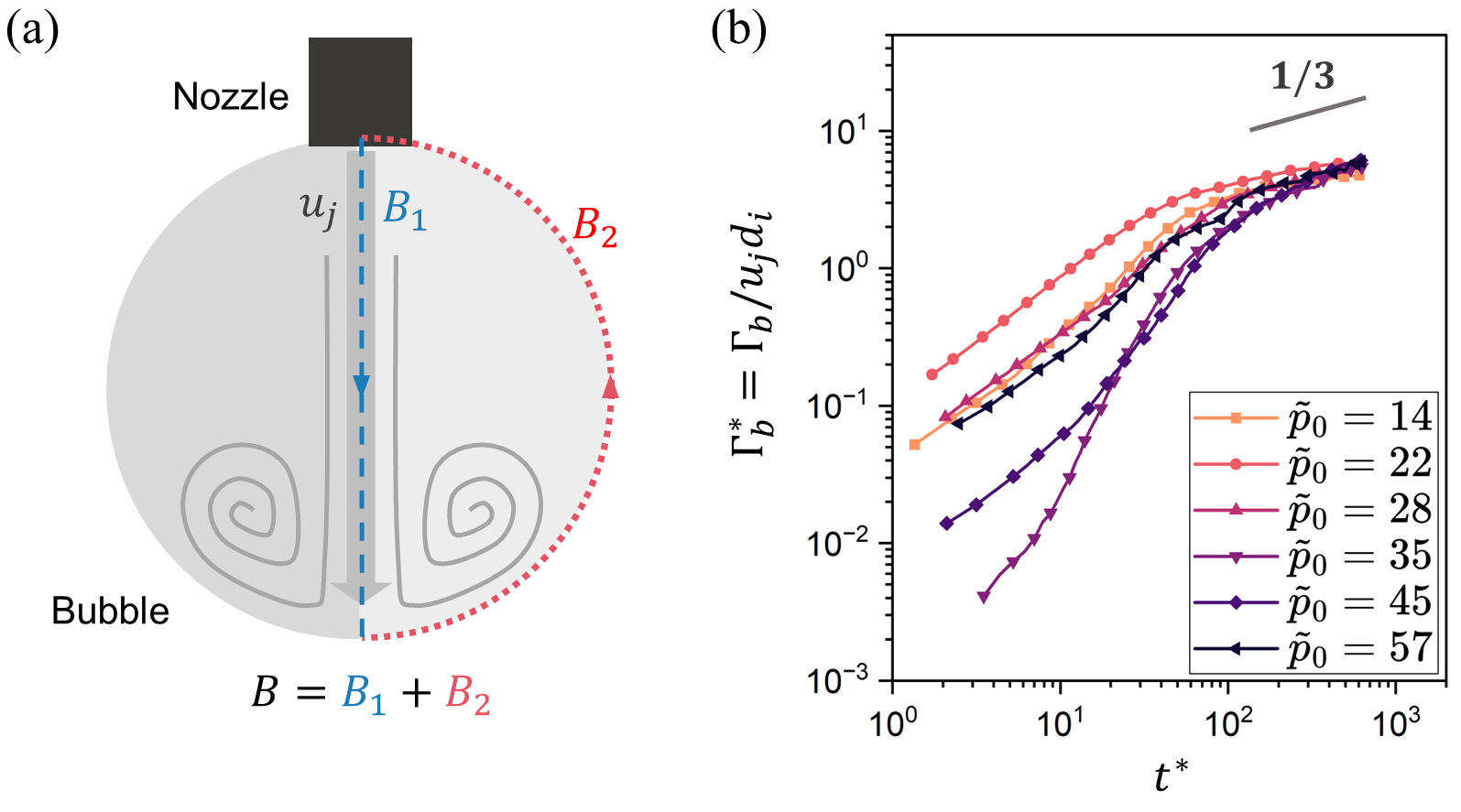}
\caption{\label{fig:bubble} (a) Contour $B$ for bubble circulation evaluation from experimental data. This contour can be segmented into $B_1$ (dashed) and $B_2$ (dotted) for simplified analysis. (b) Variation of the dimensionless bubble circulation with time depicting the scaling $\Gamma_b^\ast \sim {t^\ast}^{1/3}$}
\end{figure}

The velocity vector field obtained from PIV is analysed to understand the vortical dynamics associated with the internal flow. The circulation $\Gamma_b$ associated with the flow inside the bubble is deduced for the right half within the contour $B$ as depicted in Fig.~\ref{fig:bubble}a using $\Gamma_b=\int_{B}{\omega\ dA}$. The variation of the dimensionless circulation enclosed within bubble $\Gamma_b^\ast = \frac{\Gamma_b}{u_j d_i}$ with time $t^\ast=t u_j/d_i$ is depicted in Fig.~\ref{fig:bubble}b. We observe the scaling $\Gamma_b^\ast \sim {t^\ast}^{1/3}$ in the later stages of inflation. This can be deduced by dividing the contour $B$ into two segments $B_1$ and $B_2$ as depicted in Fig.~\ref{fig:bubble}a. Then the circulation can be also expressed as $\Gamma_b=\int_{B_1}{\textbf{u.dl}}+\int_{B_2}{\textbf{u.dl}}$, where it can be shown that $\int_{B_1}{\textbf{u.dl}} \gg \int_{B_2}{\textbf{u.dl}}$ and $\int_{B_1}{\textbf{u.dl}} \sim u_j h$ (refer to the Supplemental Material for a detailed discussion). As the height $h \sim t^{1/3}$ when the bubble is large, this predicts the observed scaling $\Gamma_b \sim t^{1/3}$.

\begin{figure}[h]
\includegraphics[width=1\linewidth]{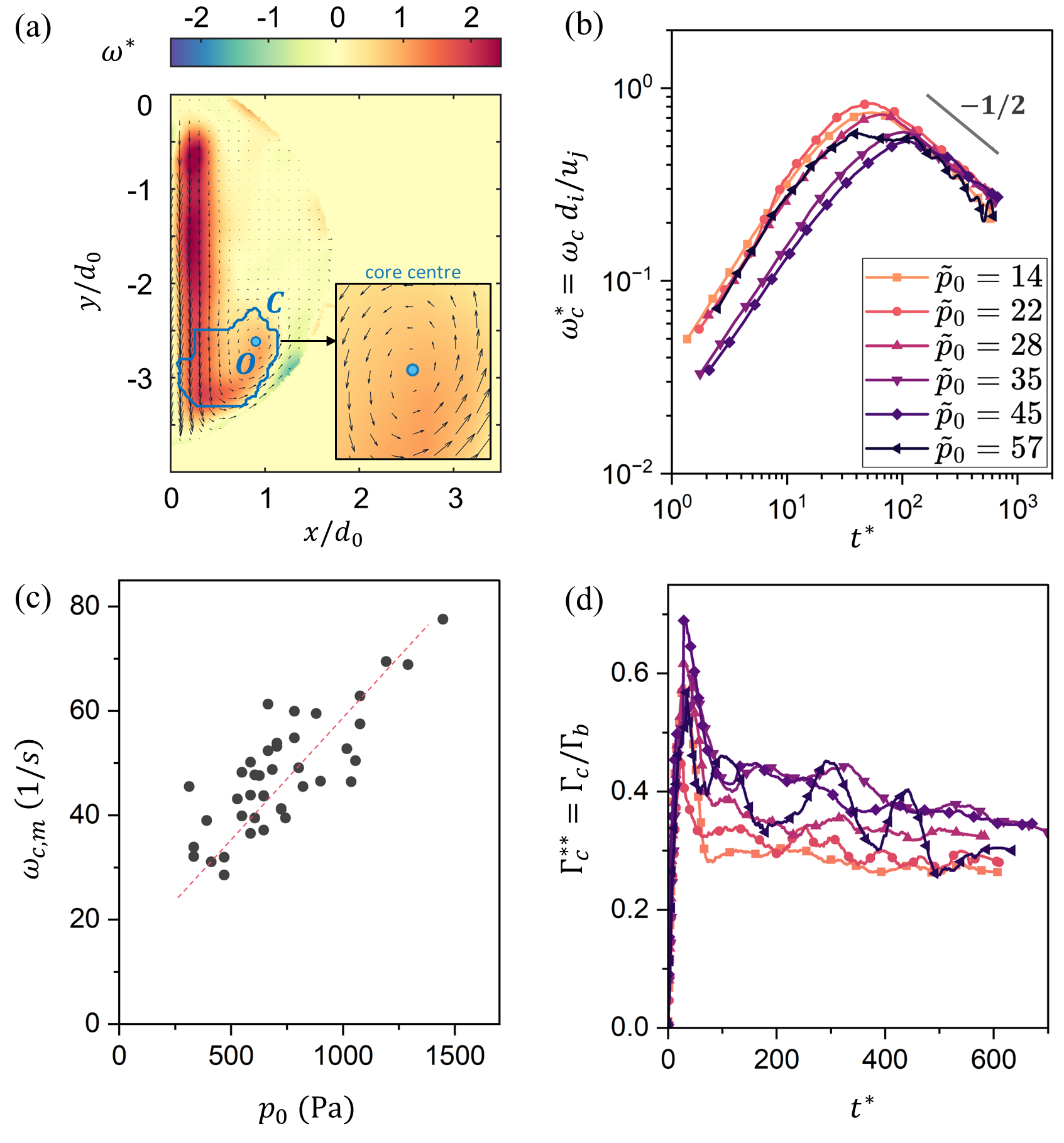}
\caption{\label{fig:core} (a) The contour \textcolor{blue}{$C$} depicting the connected region identified as core using the swirl strength ($\lambda_{ci}$) criteria with core centre \textcolor{blue}{$O$} determined from the $\Gamma_2$ method. (b) Variation of the dimensionless average core vorticity $\omega_c^\ast$ with normalised time $t^\ast$ depicting the scaling $\omega_c^\ast \sim {t^\ast}^{-1/2}$ (c) Peaks in the variation of average core vorticity $\omega_{c,m} $ correlated linearly with the source pressure $p_0$ (d) Variation of dimensionless core circulation ${\Gamma_c}^{\ast\ast}$ with normalised time $t^\ast$ depicting it approaches a constant value}
\end{figure}

The incoming jet from the nozzle has a dominant shear layer with large vorticity, contributing significantly to the bubble circulation evaluated earlier. In general, a large vorticity value does not necessarily represent a vortex core region \cite{jeong_identification_1995}. Therefore, the overall bubble circulation may not adequately describe the dynamics associated with the core. The evolution of coherent structures like a vortex core typically governs the interaction dynamics for various flow phenomena including mixing, entrainment, heat transfer, and turbulence. Henceforth, the vortex identification schemes were implemented to isolate the core for further analysis. The $\Gamma_2$ criteria \cite{graftieaux_combining_2001} was used to identify the location of the prominent core $O$, and the associated core region $C$ was determined using swirl strength or $\lambda_{ci}$ criteria \cite{zhou_mechanisms_1999}, as depicted in Fig.~\ref{fig:core}a. The $\Gamma_2$ criteria inspects the relative orientation of velocity vectors in the neighbourhood of a point to identify the core centre, and the $\lambda_{ci}$ criteria determines the core region based on the imaginary part of the eigenvalues of the velocity gradient tensor. A brief discussion of these techniques can be found in the Supplemental material. The core dynamics is assessed through the average vorticity, defined as $\omega_c=\int_{C}{\omega\ dA} / \int_{C}{dA}$ and circulation $\Gamma_c=\int_{C}{\omega\ dA}$. The variation of normalised average core vorticity $\omega_c^\ast = \frac{\omega_c}{u_j/d_i}$ with time $t^\ast=t u_j/d_i$ is depicted in Fig.~\ref{fig:core}b. We observe the scaling $\omega_c^\ast \sim {t^\ast}^{-1/2}$. A simple explanation is obtained by scaling vorticity to the velocity gradients where incoming jet velocity $u_j$ serves as a suitable scale along with a characteristic length scale associated with the core $l_c$ such that $\omega_c \sim u_j/l_c$. The identified core $C$ spans the swirling region in the neighbourhood of the core centre $O$ (depicted in the inset of Fig.~\ref{fig:core}a) to the high shear region in the vicinity of the impingement where the jet turns and spreads along the concave soap film. Although from the vector field visualisation, a vortical structure is not apparent in this impingement region, the eigenvalues of velocity gradient tensor (from $\lambda_{ci}$-criteria), however, predict a dominant rotational component. This unsteady stagnation point flow and the overall recirculatory flow associated with the wall jet separation contribute to core dynamics. Essentially, the wall jet feeds this vortex, and hence, the viscous length scale $l_c \sim \sqrt{\nu_a t}$ is suitable to characterise the vortical dynamics of the core \cite{tung_motion_1967,allen_vortex_2000}. Therefore, $\omega_c \sim u_j/\sqrt{\nu_a t}$, which explains the observed scaling $\omega_c \sim t^{-1/2}$ and the peak associated with the core vorticity variation $\omega_{c,m} \propto u_j \propto p_0$ (See Fig.~\ref{fig:core}b,c).

The core circulation is normalised using the bubble circulation such that $\Gamma_c^{\ast \ast} = \Gamma_c/\Gamma_b$. The variation is depicted in Fig.~\ref{fig:core}d, where it is observed that this dimensionless core circulation approaches a constant value with time, which means that the core circulation eventually follows a scaling law similar to the bubble circulation.

In the initial phase of inflation, core vorticity and circulation depict an increasing trend, which may be accounted for considering the flow confinement. This presumably leads to the rotational flow being concentrated in a small volume available within the bubble. Hence, a coupled growth of bubble size and vortical characteristics is observed until a point where the flow starts to separate. After this, the core vorticity starts to decay. Also, the initial phase is unsteady, with the bottom-most point of the bubble accelerating significantly. In the later stages, the jet impingement is relatively steadier. Also, the average core vorticity $\omega_c^\ast$ of this confined vortex depicts a scaling law $\sim t^{-1/2}$ independent of the bubble growth rate. The core circulation, when normalised with the overall bubble circulation $\Gamma_c^{\ast \ast}$, depicts a variation qualitatively similar to that of a free vortex ring formation \cite{gharib_universal_1998, rosenfeld_circulation_1998}. In earlier studies, this ratio approached a value of $0.4-0.6$ for free laminar vortex rings, where the total circulation of the half vortex, including the trailing jet, was used for normalisation. This is similar to what is observed in the current study. Thus, in a sense, the bubble-confined unsteady vortex ring under consideration here also retains these universal characteristics associated with the generation of a free laminar vortex ring. Although the primary source of vorticity generation is the boundary layer at the inner tube walls in the former case, and the wall jet on the bubble interface acts as an additional source in the latter. To have a better understanding of the universality of the vortical dynamics, further research is required into the unsteady phase where the wall jet separation starts and sustains.

Furthermore, the jet impingement here differs from the conventional free jet impingement \cite{bergthorson_impinging_2005, aillaud_investigation_2017} or vortex impingement \cite{ahmed_experimental_2023} studies in terms of the confinement imposed by the bubble interface. The consequences include the imposed entrainment and the vortex interacting with the jet itself. The jet eventually becomes unstable as these effects become prominent for a higher Reynolds number jet. This phenomenon is depicted in the Supplemental material.
 
\vspace{3mm}
\textbf{Conclusion:}
In closing, we have discussed the dynamics associated with an inflating soap bubble and the associated internal flow, where we deduce several universal scaling laws associated with the bubble growth and the prominent vortex that forms within. The bubble volume is found to grow linearly in time with the bubble height following the scaling ${\widetilde{h}}^3 \sim \widetilde{p}_0 \widetilde{t}$. The bubble circulation follows the scaling $\Gamma_b^\ast \sim {t^\ast}^{1/3}$ and the average core vorticity depicts $\omega_c^\ast \sim {t^\ast}^{-1/2}$. 
The phenomenon occurs prominently due to the round jet impingement with the bubble interface and was investigated within the range $Re_j=50-500$. The universality extending from a free vortex ring formation, vortical dynamics of the initial phase and the jet instability induced in the later phases are among the compelling observations that are to be examined further. 

\vspace{3mm}
\textbf{Acknowledgments:}
S. J. Rao and S. Jain would like to thank the Prime Minister Research Fellowship (PMRF) for the financial support. S. Basu would like to thank the support from the Pratt and Whitney Chair professorship.

\bibliography{apssamp}


\clearpage
\newpage

\section*{Supplemental Material}

\textbf{Abstract}: In this section, we deduce the expression for the circulation associated with the flow enclosed within the bubble. We also briefly discuss the vortex core identification schemes implemented in this study, the instability in the jet observed for higher Reynolds numbers and describe the movies provided in the Supplemental Material.

\begin{figure}[h]
\includegraphics[width=1\linewidth]{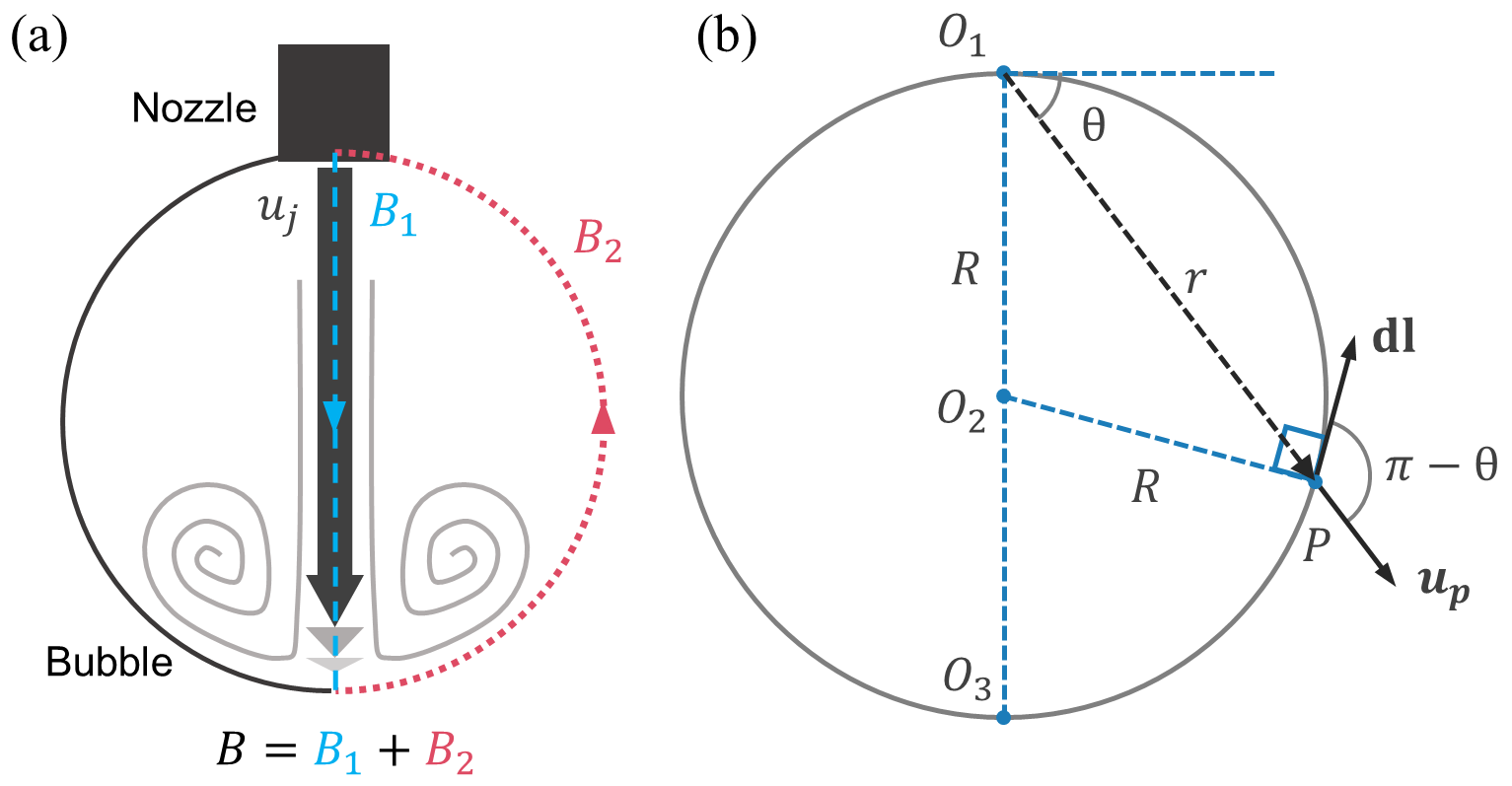}
\caption{\label{fig:Sbubble} (a) The contour $B$ for bubble circulation evaluation from experiments. This contour is segmented into $B_1$ (dashed) and $B_2$ (dotted) for simplified analysis. (b) Geometrical parameters to describe the bubble interface (arc $B_2$) and its motion in polar coordinates with origin $O_1$}
\end{figure}

\vspace{4mm}
\textbf{Derivation for Bubble circulation}: Here, we deduce the expression for the temporal variation of the bubble circulation. The circulation $\Gamma_b$ for the flow inside the bubble is evaluated from experimental data within the contour $B$ as depicted in  Fig.~\ref{fig:Sbubble}a using $\Gamma_b=\int_{B}{\omega\ dA}$. The circulation can also be expressed as 

\begin{equation}  \label{eq:circulation}
   \Gamma_b=\int_{B}{\textbf{u.dl}}=\int_{B_1}{\textbf{u.dl}}+\int_{B_2}{\textbf{u.dl}} =\Gamma_{b_1} + \Gamma_{b_2}
\end{equation}

where the contour $B$ is segmented into two parts, the diametrical chord aligned with the jet $B_1$ and the semicircle along the bubble interface $B_2$ as depicted in Fig.~\ref{fig:Sbubble}a.  

For contribution from $B_1$, a simple argument can presented that $\Gamma_{b_1} \sim u_j h$. When the bubble is large, the portion on this segment over which the stagnation occurs at the bottom will be much smaller than the bubble diameter. Hence, we can assume the scaling $\int_{B_1}{\textbf{u.dl}} \sim u_j (2R) \sim u_j h$ holds.

For contribution from $B_2$, we'll perform the integral along the arc, equivalently the bubble interface as depicted in Fig.~\ref{fig:Sbubble}b. With $O_1$ as the origin, a point on the bubble surface, say $P$, can be expressed in polar coordinates as
\begin{equation} \label{eq:interface}
    r = 2 R \sin{\theta}
\end{equation}
where the bubble radius R increases with time and can be expressed as $R = \widehat{R} t^{1/3}$ as deduced earlier from the large bubble approximation, where $\widehat{R}$ is constant for a given source pressure. Furthermore, the gas phase velocity inside the bubble at the interface should match the boundary velocity. Therefore, the fluid velocity in the radial direction at this arbitrary point $P$ (using Eq.~\ref{eq:interface}) is 

\begin{equation} \label{eq:interface_vel}
    u_p = \frac{dr}{dt} = \frac{2}{3}\widehat{R}t^{-2/3}
\end{equation}

Substituting this to get $\Gamma_{b_2}$ along with $dl = r d\theta$ gives

\begin{equation} \label{eq:interface_vel}
    \Gamma_{b_2} = \int_{\pi/2}^0 {u_p \space dl \cos{(\pi-\theta)}} = \frac{4}{9}{\widehat{R}}^2 t^{-1/3} = \frac{4R^2}{9t}
\end{equation}

The relative strength of the circulation contributed from $B_1$ and $B_2$ is hence 

\begin{equation} \label{eq:circulation_rel}
    \frac{\Gamma_{b_1}}{\Gamma_{b_2}} \sim 9 \frac{u_j}{2R/t} \sim 3 \frac{u_j}{u_{o_3}}
\end{equation}

where this is essentially the ratio of incoming jet velocity to the bubble interface velocity $u_{o_3}$ at the bottommost point. The bubble interface grows slowly when the bubble is large, and hence, for this impingement kind of interaction between the jet and bubble, we have  $u_j \gg u_{o_3}$. From Eq.~\ref{eq:circulation_rel}, we have $\Gamma_{b_1} \gg \Gamma_{b_2}$. The contribution from $B_2$ is therefore insignificant in the later stages of inflation and hence $\Gamma_{b} \approx \Gamma_{b_1} \sim u_j h$. As the bubble height $h \sim t^{1/3}$ in this regime, this predicts the observed scaling $\Gamma_b \sim t^{1/3}$.

\vspace{4mm}
\textbf{Vortex core identification schemes}: 
 Detecting vortices using vorticity fields may lead to false estimates \cite{jeong_identification_1995} as they fail to differentiate between shear layers and the desired rotationally dominant regions. In this study, amongst the various available methods, we implement two schemes in conjunction, namely the $\Gamma_2$ criteria \cite{graftieaux_combining_2001} and the swirl strength or $\lambda_{ci}$ criteria \cite{zhou_mechanisms_1999} to locate the core within the bubble from the two-dimensional PIV vector field. $\Gamma_2$ criteria is a robust method to detect the core centre based on the relative orientations of the velocity vectors within a prescribed neighbourhood of the point under consideration. The $\Gamma_2$ parameter is defined as

\begin{equation}   \label{eq:gamma2}
\Gamma_2(P) = \frac{1}{N}\sum_S \frac {({\textbf{PQ}} \times (\textbf{U}_Q - \textbf{U}_m))\cdot \textbf{z}}{\parallel \textbf{PQ} \parallel\cdot \parallel \textbf{U}_Q - \textbf{U}_m \parallel}
\end{equation}

which is determined at each point $P$ in the vector field. $\textbf{PQ}$ denotes the displacement vector from point $P$ to $Q$ inside area $S$ in the neighbourhood of $P$ containing $N$ points, $\textbf{U}_m$ refers to the mean velocity of the fluid enclosed within $S$, $\textbf{U}_Q$ refers to the fluid velocity vector at point $Q$ and $\textbf{z}$ is the unit vector normal to the plane. The peak in the $\Gamma_2$-field corresponds to the core centre location.

The swirl strength criteria \cite{zhou_mechanisms_1999} determine the vortex or regions with the rotationally dominant flow by considering the eigenvalues of the velocity gradient tensor. The imaginary part of the complex eigenvalue pair is the local swirling strength or $\lambda_{ci}$. A vortex is then a connected region where the $\lambda_{ci}$ value is positive. 

The vortex core of interest in the present study is the connected region identified by the $\lambda_{ci}$ criteria, enclosing the dominant core centre identified using the $\Gamma_2$ method (with neighbourhood $S: 2.1 \times 2.1$ mm)

\vspace{4mm}
\textbf{Jet instability}: 
We observed that the bubble inflation displays a round jet feature impinging with the soap film concave interior. The jet impingement within this enclosed confinement includes consequences like forced entrainment and self-interaction with the vortical structure. These effects become prominent as the bubble grows in size for a higher Reynolds number jet and it eventually becomes unstable. The region of interest is shifted downwards to observe this phenomenon, and the captured flow structures are depicted in Fig.~\ref{fig:unstable} for a stable and unstable jet case (see supplemental movies 2-3). In the unstable jet case, i.e. Fig.~\ref{fig:unstable}b, we observe small vortical structures embedded within the wall jet and accumulating into the larger coherent vortical structure. The transition is found to happen roughly beyond $Re_j \approx 250$ for the bubble heights within $\widetilde{h} \approx 10$ in the current study. Additional investigation is required to comprehend this transition, presumably driven by a feedback mechanism that occurs even within this low Reynolds number regime. 

\begin{figure}[h]
\includegraphics[width=1\linewidth]{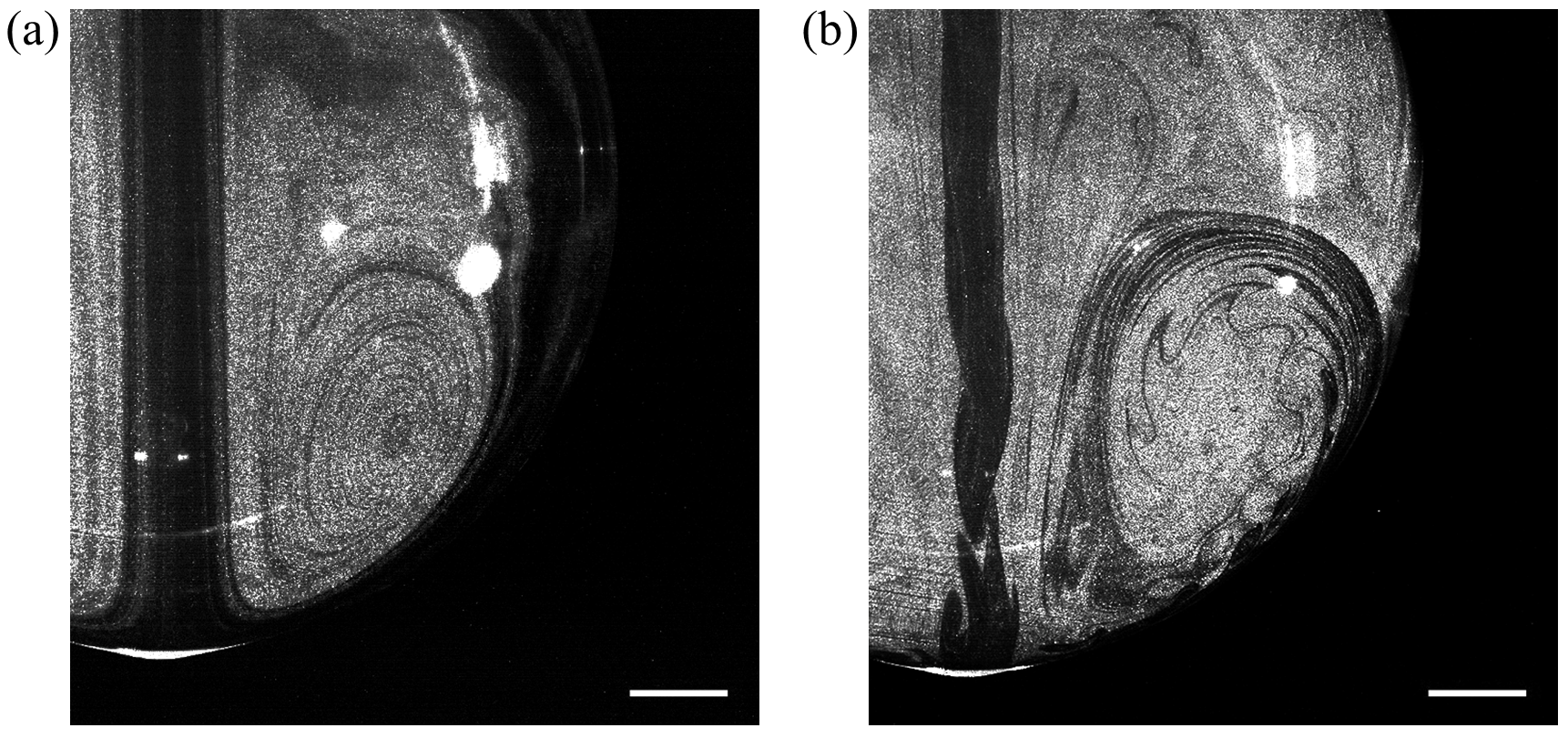}
\caption{\label{fig:unstable} The jet morphology at later stages of inflation captured at a field of view $30mm$ below the nozzle tip. The scale bar represents $10mm$ (a) stable jet - $Re_j = 172$ at $t^\ast = 2362$ (b) unstable jet - $Re_j = 481$ at $t^\ast = 2821$}
\end{figure}

\vspace{4mm}
\noindent \textbf{Movie Captions:}

\begin{enumerate}

    \item Visualisation of the internal flow with ROI near the nozzle for the normalised source pressure $\widetilde{p}_0 = 26.54$ and jet Reynolds number $Re_j \approx 150$. 

    \item Visualisation of the internal airflow during bubble inflation with ROI $30 mm$ below the nozzle for the normalised source pressure $\widetilde{p}_0 = 30.32$ and jet Reynolds number $Re_j \approx 172$. A stable jet is observed.

    \item Visualisation of the internal airflow during bubble inflation with ROI $30 mm$ below the nozzle for the normalised source pressure $\widetilde{p}_0 = 84.66$ and jet Reynolds number $Re_j \approx 481$. An unstable jet is observed.

    \item A larger field-of-view visualisation of the internal airflow during bubble inflation for the normalised source pressure $\widetilde{p}_0 = 17.12$ and jet Reynolds number $Re_j \approx 115$.

    \item A larger field-of-view visualisation of the internal airflow during bubble inflation for the normalised source pressure $\widetilde{p}_0 = 26.54$ and jet Reynolds number $Re_j \approx 150$.

    \item A larger field-of-view visualisation of the internal airflow during bubble inflation for the normalised source pressure $\widetilde{p}_0 = 82.19$ and jet Reynolds number $Re_j \approx 467$.
    
\end{enumerate}

Note: A planar laser sheet illuminates the airflow seeded with olive oil seeder particles. For scale, the nozzle outer diameter is $10 mm$. Movies 1-3 were recorded with the Photron Mini UX high-speed camera, synchronised with the pulsed laser source. Movies 4-6 were recorded using a Nikon 7200 DSLR camera, where a moving black strip can be observed as the camera was synchronised with the laser. All videos run at actual speed.


\end{document}